# Kinesthetic Learning – Haptic User Interfaces for Gyroscopic Precession Simulation

Felix G. Hamza-Lup

Computer Science
Georgia Southern University
Savannah, GA 31419, US
*E-mail: fhamzalup@georgiasouthern.edu*

**Abstract.** Some forces in nature are difficult to comprehend due to their non-intuitive and abstract nature. Forces driving gyroscopic precession are invisible, yet their effect is very important in a variety of applications, from space navigation to motion tracking. Current technological advancements in haptic interfaces, enables development of revolutionary user interfaces, combining multiple modalities: tactile, visual and auditory. Tactile augmented user interfaces have been deployed in a variety of areas, from surgical training to elementary education. This research provides an overview of haptic user interfaces in higher education, and presents the development and assessment of a haptic-user interface that supports the learner's understanding of gyroscopic precession forces. The visual-haptic simulator proposed, is one module from a series of simulators targeted at complex concept representation, using multimodal user interfaces. Various higher education domains, from classical physics to mechanical engineering, will benefit from the mainstream adoption of multimodal interfaces for hands-on training and content delivery. Experimental results are promising, and underline the valuable impact that haptic user interfaces have on enabling abstract concepts understanding, through kinesthetic learning and hands-on practice.

**Keywords**: Haptics, Force Feedback, Gyroscope, Precession, Computer-based Simulation.

## 1. Introduction

Torque-induced precession (i.e., gyroscopic precession) is a physical phenomenon, in which the axis of a spinning object (e.g., a gyroscope) describes a cone in space when an external torque is applied on it. One can *feel* the precession forces by spinning a wheel, and attempting to modify the spinning axis orientation. Gyroscopes serve a very important function in both simple and highly advanced navigational devices, because precession and angular velocity are integral to modern navigation concepts. From air to sea, these concepts help pilots determine height, depth and various other pieces of information required for safe navigation. Gyroscopes come in a wide variety of forms, from mechanical to optical gyroscopes, from macro to micro-scale, and they are employed in systems for guidance, attitude reference and stabilization, applications for tracking and pointing, as well as flight data analysis (Passaro et al, 2017). Understanding the relationship among gyroscopic precession, angular momentum and angular velocity is a fundamental part of college level physics and engineering education worldwide. Gyroscopic precession, conservation of momentum and other associated abstract concepts, are difficult to understand by freshmen, and faulty mental models can generate confusion in their minds. Many students have difficulty understanding abstract physics and/or mechanical engineering concepts taught using traditional teaching methods. When learners resort to memorization rather than reasoning, they will find it difficult to apply and adapt what they learn to new situations. In the US, the Science, Technology, Engineering and Mathematics (STEM) initiative is targeted at helping learners gain knowledge and hone their reasoning skills. Individual experimentation and observation of force vectors, as well as the simulation of abstract concepts, facilitates and improves the learners' mental models and capacity to understand complex systems. Understanding complex systems and holistic thinking, is an essential skill for engineers (Nelson et al, 2010). Spatial visualization skills and correct judgement of forces are fundamental to a variety of disciplines, but are particularly important for STEM disciplines (Uttal and Cohen, 2012).

Haptic (e.g., *force-feedback* or *vibro-tactile*) interfaces have been increasingly used over the past decade to convey tactile information through Haptic-based User Interfaces (HUI). From early stages of education, humans learn to identify various objects and concepts through the sense of touch, as kinesthetic learners hence, it makes sense to augment the visual channel provided by a Graphical User Interface (GUI) with tactile components. Using multimodal interfaces to present abstract concepts, has the potential to increase the learner's engagement and his understanding capacity. In an effort to improve abstract concept delivery to learners, we propose a haptic enhanced user interface for the simulation of the forces involved in the gyroscopic precession. The cost-effective system was deployed and assessed in a laboratory setup, with the help of a sizable group of volunteers.

The paper is organized as follows: Section 2 provides an overview on related work vis-à-vis haptic interfaces for multimodal content delivery, with an emphasis on haptic systems for simulation and training. In Section 3, the background theoretical concepts associated with Gyroscopic Precession (GP) are presented. Section 4 describes the implementation of the visual-haptic simulator, beginning with the motivation and the goals for the simulator development, followed by the description of the graphical and the haptic user interfaces. Section 5 defines the experimental setup and the participants partitioning. In Section 6, the assessment methodology is presented, and the analysis of the experimental results, followed by the conclusion and closing remarks.

## 2. Haptic User Interfaces for Simulation and Training

Haptic User Interfaces provide users with cutaneous feedback and/or kinesthetic/force-feedback during interaction with computer generated virtual elements or remote objects manipulation (robotic tele-manipulation). Haptic devices come in a wide variety of forms and shapes, from vibro-tactile systems, to complex robotic arms that track the position and orientation of the user's arms.

### 2.1. Haptic Technology Drivers

Haptic systems development is primarily driven by the medical field (i.e., surgical simulators, complex medical procedures) and the entertainment industry (i.e., video gaming). In the video gaming industry, HUIs have been heavily employed to increase realism by adding the sense of touch. Game development companies (e.g., Electronic Arts) invested heavily in the technology that develops haptic controllers to bring enhanced realism into gaming through "real-pain" sensations (Stone, 2018). Popular games, such as Half-Life 2, support the use of the Novint Falcon (Novint, 2018) haptic devices with a "pistol grip" accessory.

However, even before the spread of haptic systems in the video gaming industry, the touch modality was explored in medical simulation and training. Medical procedures education with haptic feedback provides many advantages for training (Hamza-Lup et al, 2011) and, over the past decade, several research and industrial-level efforts, lead to a set of APIs and software frameworks for haptic feedback integration into existing user interfaces (Popovici et al, 2012). Along the same direction, rehabilitation and disability services are very well suited for haptic-based user interfaces.

For example, recently (Bortone et al, 2018) proposed a wearable haptic systems for rehabilitation of children with neuro-motor impairments. Many other examples and prototypes have been proposed, for a survey on this topic please see (Newton et al, 2019).

## 2.2 Haptic Technology in Education

The successful application of haptic user interfaces in education is based on two fundamental principles:

1. *Hands-on learning*, empowering kinesthetic and tactile learners, allowing the learner to experience, manipulate, and understand through first-hand interaction. Such learners have characteristics that facilitate their learning through touch (Child1st, 2018).
2. *Gamification*, i.e., using game mechanics and methods in teaching contexts to increase the learner's engagement, participation, and competition (Kim et al, 2018).

Early experiments in education (Jones et al, 2005), proves that touch gives learners a feeling of being more involved in learning, and an increased connection with the learning material. The haptic paradigm applied in education overcame many challenges in recent years, and many prototypes have been proposed in conjunction with 3D user interfaces (Hamza-Lup and Stanescu, 2010). When learners use the haptic interface they become more interested in the material as compared to individuals who learn only through traditional methods (Weibe et al, 2009). However, proper introduction to such interfaces must be completed, in order to cope with the additional cognitive demands on the user's side.

Haptic user interfaces have been proposed to aid in the understanding abstract concepts in physics: e.g., friction coefficients and forces on an inclined plane (Hamza-Lup and Baird, 2012), engineering dynamics, inertia (Okamura et al, 2002). Mechanical concepts simulations using haptic augmentation have been proposed for understanding pulley systems and the linear acceleration increase based on the radius of the pulley (Neri et al, 2018). Using a 2D visual component and a haptic device, the user can virtually pull on a string attached to a pulley system, and feel the forces acting on the string. Moreover, the pressure model, and its dependency on force amount and surface area, is essential for any engineer that works with hydraulic systems. The Haptek16 (Hamza-Lup and Adams, 2008), uses force feedback systems to enable learners to experiment and gain a deeper understanding of hydraulics concepts. Many other prototype HUIs exist, however, very few provide a comprehensive assessment, and none have tackled the simulation of complex and abstract gyroscopic precession pseudo-forces.

## 3. Gyroscopic Precession

Gyroscopes are very useful in navigation, especially where magnetic compasses do not work, such as in manned and unmanned spacecraft, ballistic missiles, unmanned aerial vehicles, and satellites (e.g., space telescopes). Gyroscope associated paradigms are proposed for the generation of alternative "gravitational like" forces through gyration in futuristic NASA space exploration prototypes.

A gyroscope can be defined as a spinning disk, in which the axis of rotation is allowed to assume any orientation. When spinning the rotor, the orientation of the spin axis is not affected by the orientation of the body that encloses it, and the body enclosing the gyroscope can be moved in space without affecting the orientation of the spin axis as illustrated in Figure 1.

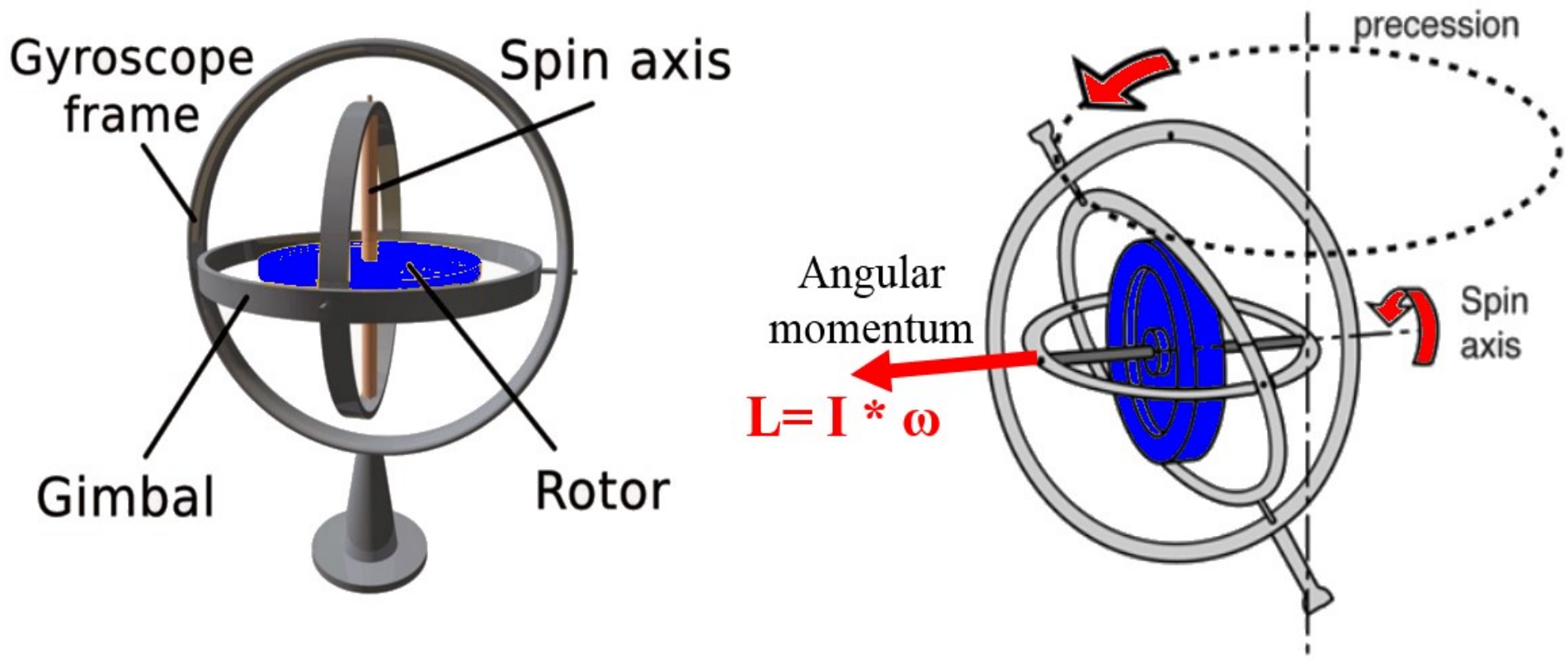

Figure 1. Gyroscope components and gyroscopic precession

Precession is the change of angular velocity and angular momentum produced by a torque. The torque is a measure of how quickly an external force can change an object's angular momentum, either magnitude, direction, or both. As angular momentum decreases, gravitational forces cause the end of the axle to precess in subsequently smaller circles (as illustrated in Figure 1). The angular momentum equation is given by:

$$L = I\omega\ ,$$

where $I$ is the inertia (dependent on the object's mass) and $\omega$ is the angular velocity. To find the direction of the angular momentum, one must use the Right-Hand rule. The general equation that relays the torque to the rate of change of the angular momentum is:

$$\tau = \frac{dL}{dt},$$

where $\tau$ is the torque and $L$ is the angular momentum. Moreover

$$\tau = rMg\sin\Theta,$$

where $r$ is the distance between the spin axis's origin and the force handle, $M$ is the mass of the object, $g$ the gravitational constant and $\Theta$ the precession angle.

For most learners, an easy way to understand why gyroscopic precession occurs, without using any mathematics, is looking at the behavior of a spinning object. For example, a spinning wheel possesses a property known as *rigidity in space*, meaning that the spin axis resists any change in orientation. However, gyroscopic precession is generally more perplexing to learners than the two-dimensional problems considered in an introductory physics or mechanics courses. The fact that it is more difficult to rotate a spinning wheel than a stationary one (as illustrated in Figure 2), is non-intuitive, and the direction of the force exerted by the axle on the person is unexpected.

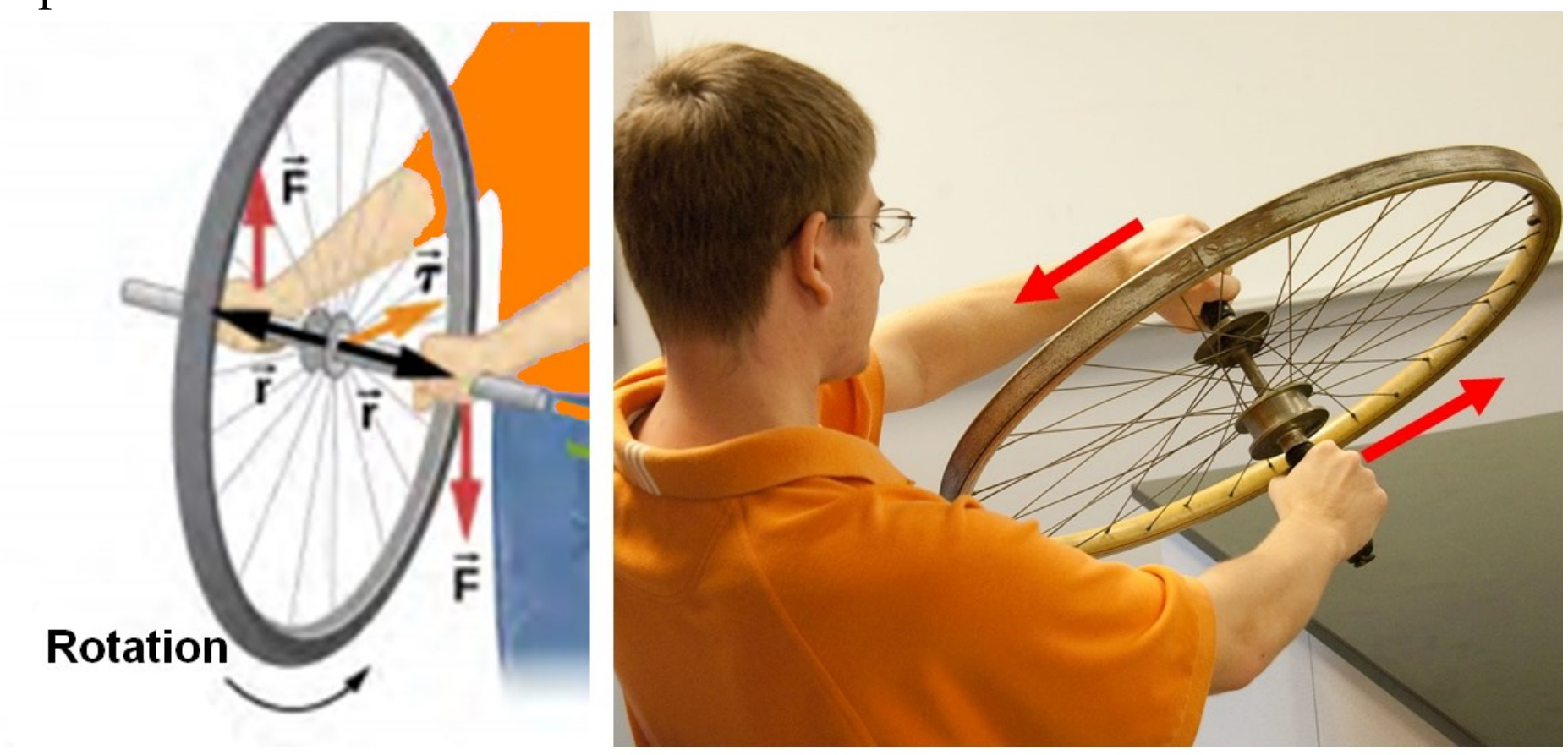

Figure 2. Feel the gyroscopic precession forces tilting a spinning wheel.

In higher education, gyroscopic precession and associated concepts (e.g., nutation) are very important concepts and building blocks for mechanics engineering and physics students. Mechanical engineering students and physics majors must possess a clear mental model and a thorough understanding of these concepts. A number of visual simulations have been proposed (Butikov, 2006), and can aid in the abstracts concepts understanding, however, none provide a realistic 3D accurate representation of these phenomena with force-feedback support.

# 4. Gyroscopic Precession Simulator Implementation

Gyroscopic precession explanations in physics textbooks are highly mathematical and hard to understand conceptually. The invisible and non-intuitive forces generated in gyroscopic precession, as well as the 3D aspect of vector composition and Right-Hand rule applications, make gyroscopic precession a difficult concept to understand by many students. Memorization and the lack of individual independent experimentation makes students unfit for engineering industry challenges (Nieh et al, 2018).

## 4.1 Motivation and Goals

Spinning the bicycle wheel scenario is an excellent way to feel the forces, and individually experiment. However, it is unrealistic when twenty-five or more students must hold and spin bicycle wheels during laboratory hours (one can only imagine the potentially dangerous outcome of such an event). Additionally changing the parameters of the wheel (e.g., shape, mass, radius, speed) in a continuous fashion, is not possible due to the physical constraints of the experiment.

Based on previous experiments (Hamza-Lup and Page, 2012), we hypothesize that, the ability to change the simulation parameters while having simulation consistency, will allow learners to effectively get "hands on" experience from a wide range of scenarios, and they will be able to better understand how the radius of the wheel, the mass and other parameters affect the magnitude of the precession force vectors.

When abstract concepts are presented, accompanied by complex mathematical formulae, many learners' cognitive levels are overloaded, and as a consequence, their attention shifts away from the learning material. The visual-haptic simulator proposed has the following goals:

- Enhance learners' engagement during class by introducing a new modality in abstract concept presentation, taking advantage of the sense of touch and 3D visualization.
- Provide the ability to fine-tune the parameters of the simulation. A change in any variable would necessarily change the outcome of the forces being applied and felt. The learner will be in control of each variable, and therefore, in control of the simulation. Changing the angular velocity, wheel weight and the length of the handle, each user has the ability to feel the forces in a dynamic environment.
- Ensure that each learner has access to the simulation. From the hardware point of view, the low-cost trend in haptic hardware enables the proliferation of this technology in every household. Setting up an entire laboratory with such devices is becoming economically feasible. The software used to develop the interface is open source and relies on international open standards.

## 4.2 Graphical User Interface

For the graphical user interface (GUI), the X3D international standard was employed. Extended 3D (X3D, 2018) is an ISO standard that provides an extensive set of features to implement innovative user interfaces on the Web, in order to support and augment various knowledge sharing and collaborative design activities. X3D is developed by the Web3D Consortium (Web3D, 2018) as a superior Web3D standard, improving on many issues from the past (e.g., XML compliancy, format etc.).

The GUI was designed to replicate a real-world scenario of a spinning bicycle wheel. It consists of an X3D model of a bicycle wheel and several menus, that allow the adjustment of the gyroscopic precession simulation, as illustrated in Figure 3 (a). The Haptic User Interface is implemented using various haptic devices. Figure 3 (b), shows the implementation using two Falcon Novint devices.

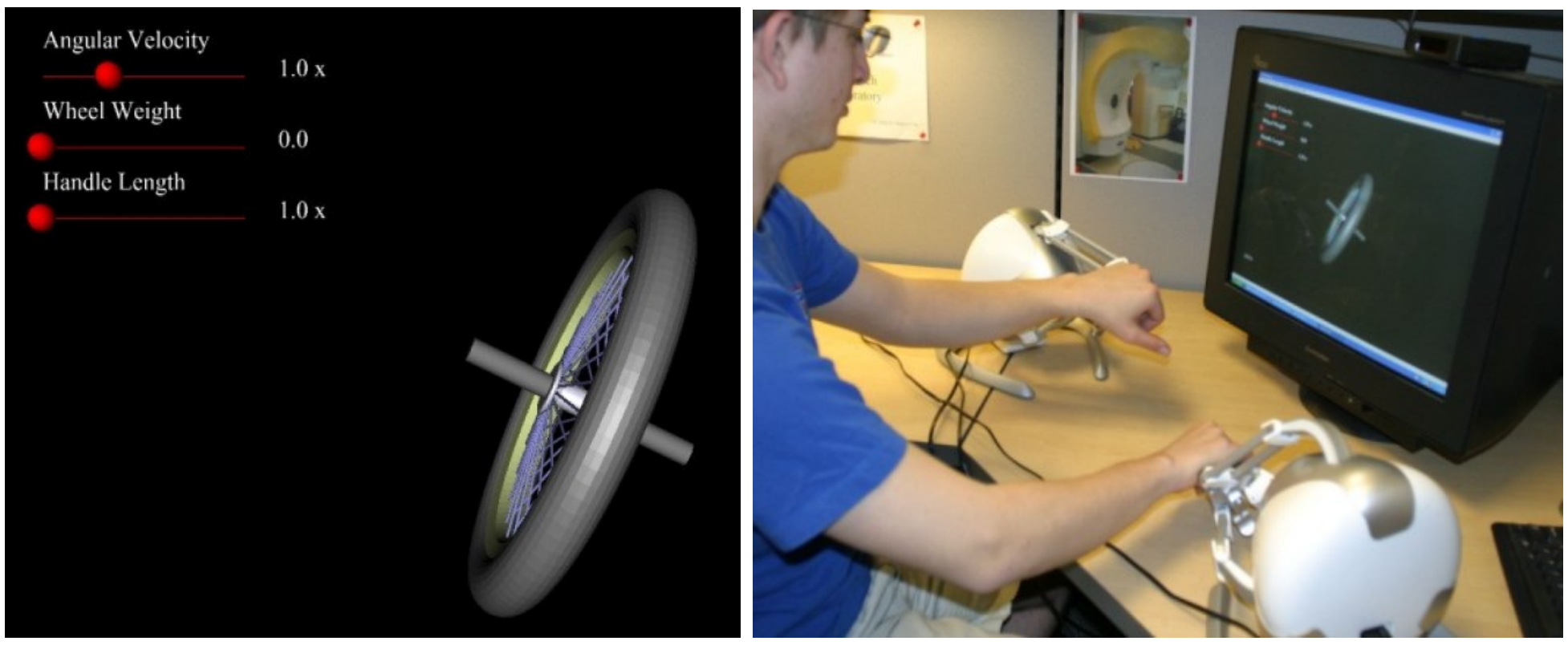

(a) (b)

Figure 3. (a) 3D GUI components, (b) HUI, with Falcon Novint devices

The user can change various simulation parameters, and has a 3D view of the rotating wheel using various 3D display technologies e.g., inexpensive reb/blue filter glasses or Crystal-Eyes shutter glasses (Crystal-Eyes, 2018).

## 4.3 Haptic User Interface

User's force-feedback is provided through two Falcon Novint devices (as illustrated in Figure 3 b). As the user is changing the position of the spinning axis, he will feel a counter-force (i.e., the precession pseudo-forces) in the Falcon's handle.

The counter force magnitude has been experimentally adjusted to provide the same "feeling" as a real 23 inch inner-diameter metal spinning bicycle wheel (as illustrated in Fig. 4). The reason for this force similarity is that, in the current laboratory experiments, the instructor uses such a wheel to demonstrate the concept and the direction of forces. However the tradition setup does not allow all students to spin wheels during lab time to practically experience the force vectors. Such experiments are now possible employing the proposed system.

The HUI was implemented using the H3D (H3D, 2018), a haptic API dedicated to haptic modeling that combines the OpenGL and the X3D standards with haptic rendering in a single scene graph, merging haptic and graphic components. H3D is independent of the haptic device, it is multi-platform, and allows audio, as well as 3D stereoscopic device integration.

H3D is conceived to support rapid prototyping, combining the power of C++ and Python scripting to improve the speed of execution.

The force-feedback effects are generated using the H3D's *PositionFunctionEffect* node. The node creates a force in the scene controlled by a position function defined on each of the X, Y and Z directions. The spherical constraints for motion are attained by calibrating each haptic device X and Y coordinates in relation to each other.

To compensate for the servomotors' slight vibration, a *SpringEffect* node is added for each device. The node generates a localized haptic effect, where, the haptics device is pulled towards to a point in space in a spring like manner. The dampening effect of the spring will compensate for the servomotors' slight vibration, pretty much like a car suspension system attenuates road vibrations.

Gyroscopic precession pseudo-forces are simulated using the H3D's *ViscosityEffec*t node that specifies a force in the opposite direction of the movement of the haptic device. The viscosity effect matches perfectly, tactile wise, the feeling given by the precession triggered forces that resist changing the spin axis's orientation.

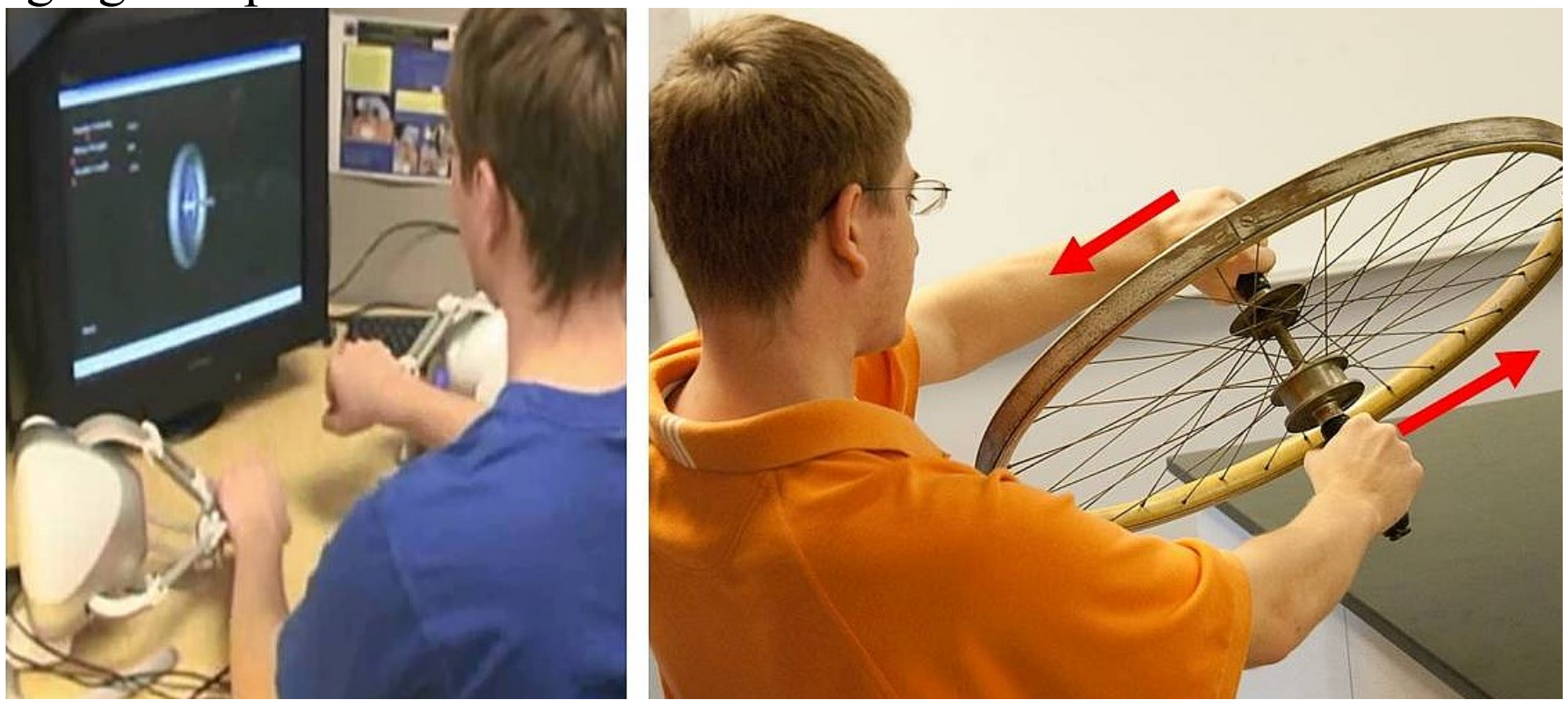

Figure 4. User changes the virtual spinning axis and feels counter-forces

To ensure the haptic devices follow a spherical path, and to synchronize the two devices and simulate the feeling of holding a single spin axis (as a rigid bar, illustrated in Figure 4), we mapped the two devices in a negative relationship to each other (i.e., the paired positive coordinate *Y* would receive a negative value). The haptic process is synchronized with the visual process by computing the wheel's rotation in relation to haptic devices real world Cartesian coordinate, using a sinusoidal function.

# 5. Experimental Setup

Haptic simulator efficiency, as an addition to traditional instruction, was evaluated through the deployment of a set of laboratory activities, designed to facilitate understanding of the precession force vectors directions and magnitudes under a wide range of conditions. Participants were asked to change various parameters of the wheel, and then, experiment with the simulator and feel the variation (increase or decrease) in the force vectors magnitude, as well as observe their change in direction.

## 5.1 Experiment Participants

The simulator assessment included 64 undergraduate students with non-physics background that volunteered for the study. The participants were aged between 19 and 31 years old, with the group Mean (M) = 22.5 years, and Standard Deviation (SD) = 2.35. The participants were divided into two equal-number groups:

- The *Control group*, denoted *C*, composed of 32 participants, aged 19 to 29 (M=23.4, SD=2.5). These participants used traditional methods to learn about the gyroscopic precession in a traditional laboratory, through paper-based problems solving and experiments conducted by the teacher during the laboratory time.
- The *Visual-Haptic group*, denoted *VH*, composed of 32 participants, aged 19 to 31 (M=24.3, SD=2.3). These participants used the visual-haptic simulator during the laboratory instruction. Each participant experimented (i.e., “played”) with the visual-haptic simulator individually and independently, while experiencing through tactile sensation the force vectors’ magnitude and direction.

To ensure uniformity in terms of previous performance, the distribution of students between groups was such that the combined grade point average (AGPA) for each group was similar, with a small SD, that is the VH group AGPA was *2.96* with a SD of *0.31*, and the C group AGPA was *2.92* with a SD of *0.25*.

## 5.2 Hardware and Experimental Setup

A physics laboratory was used as the main location for deploying the haptic devices. Sixteen laptop computers with USB2 interfaces (enabling easy haptic device connection) were used to setup the simulation environment. We used, a low-cost version of Falcon Novint™ (Novint, 2018), a desktop haptic robot device, also referred to as a "*3D mouse*", employed mainly as a gaming peripheral by many videogames enthusiasts. The device consists of three motorized arms attached to an interchangeable end-effector in the standard form of a ball grip, as illustrated in Figure 5.

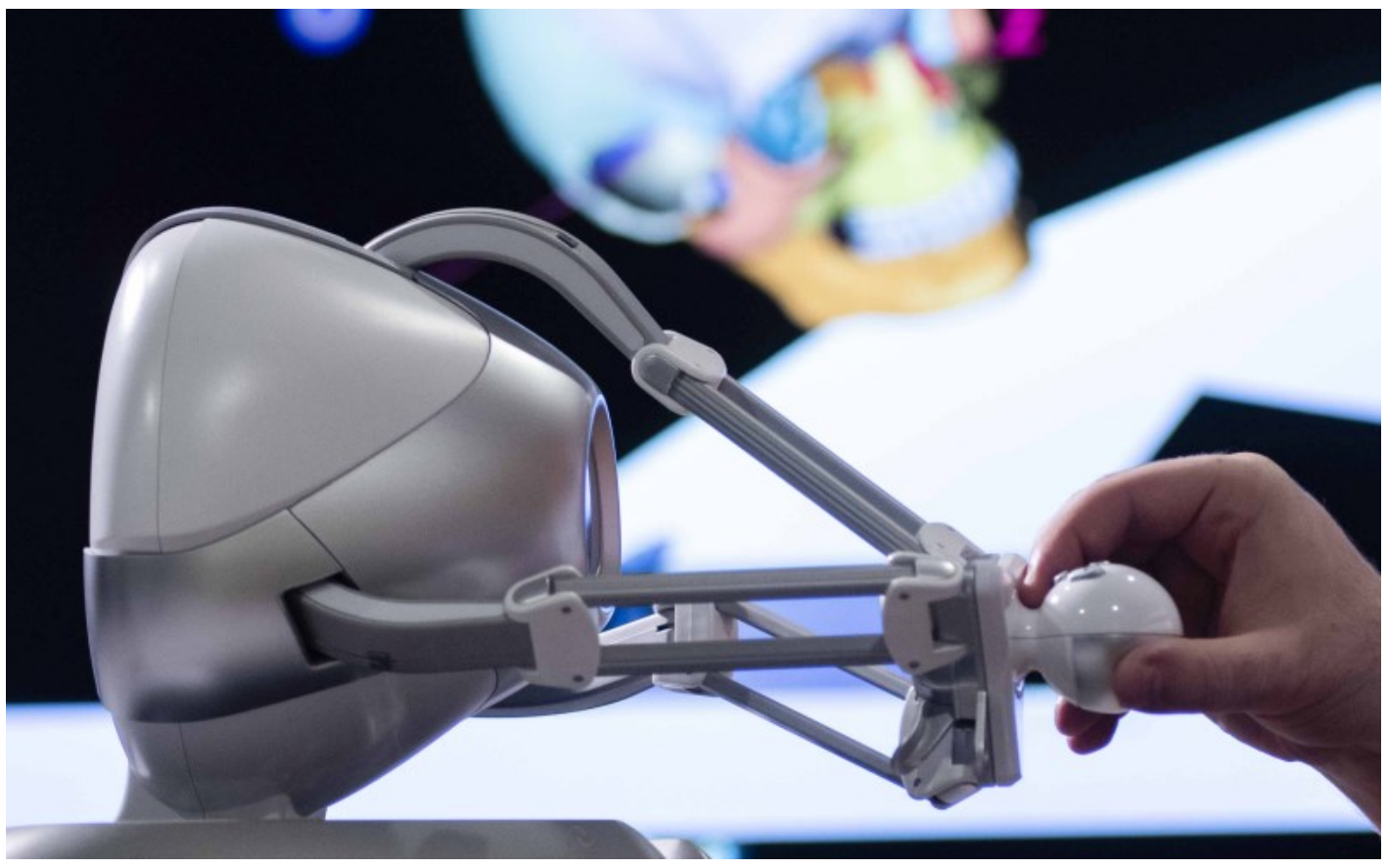

Figure 5. Falcon Novint – haptic device and ball grip attachment

The haptic device is connected to the computer system using a USB2 interface, has a 4 cubic inches 3D touch volume, and can apply up to 2 lbs. of force (i.e., approx. 9 Newtons). It can simulate the tactile feeling of objects with sub-millimeter precision, having a refresh rate of about 1 KHz. 32 devices in pairs of two were deployed in a laboratory setup, as illustrated in Figure 6.

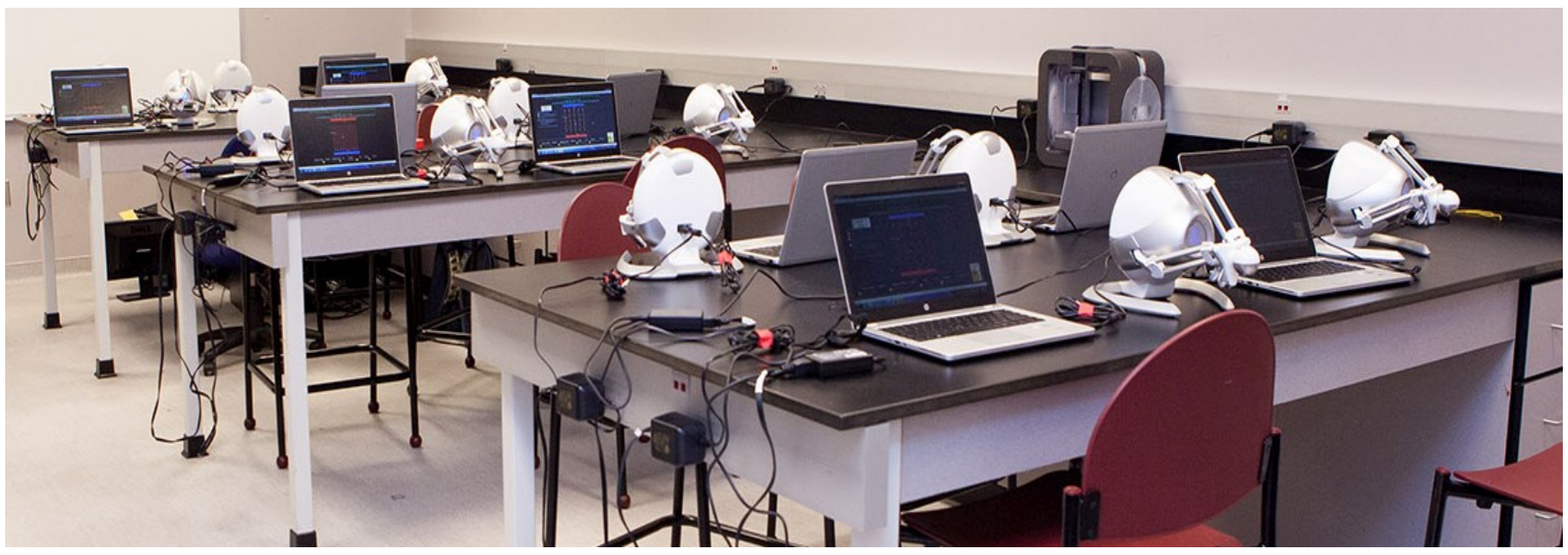

Figure 6. Haptic devices setup in the laboratory, half-room.

All the computes employed had an I5 quad-core INTEL (2.4GHz) processor with 8GB of RAM and similar hardware components (GPU, peripherals and storage). The software was installed through a mirror server on all computers, and measures were taken to assure all machines ran similar processes (e.g., all software backup, update functions, as well as internet and Bluetooth connections were turned off, etc.), prioritizing the simulation's processes.

## 6. Assessment Procedure and Experimental Results

The visual haptic simulator assessment was executed following three steps:

(1) In the first step, all participants received and completed a pre-tests (or background knowledge assessment).
(2) The second step consisted in a lecture, followed by laboratory experiments using the traditional (C group) approach, and the visual-haptic simulator (VH group) approach.
(3) The third step consisted in the post-test evaluation of all the participants. Pre and Post-test scores were collected and anonymized (i.e., student names were replaced with an id from 1 to 32 followed by the group identifier, C respectively VH, depending on the case).

An average grade, rounded up to the first two decimals was assigned for each test based on the grades provided by three independent graders. The anonymized student id was used, to reduce potential for bias, and graders were unaware of the experiments.

### 6.1 Background Knowledge Assessment – Pre-Test

Before participating in the learning activity, all the participants in the experiment took a pre-test in the form of a questionnaire of 25 equally weighted questions. The test was designed to evaluate the participants' prior knowledge of the Gyroscopic Precession and associated concepts.

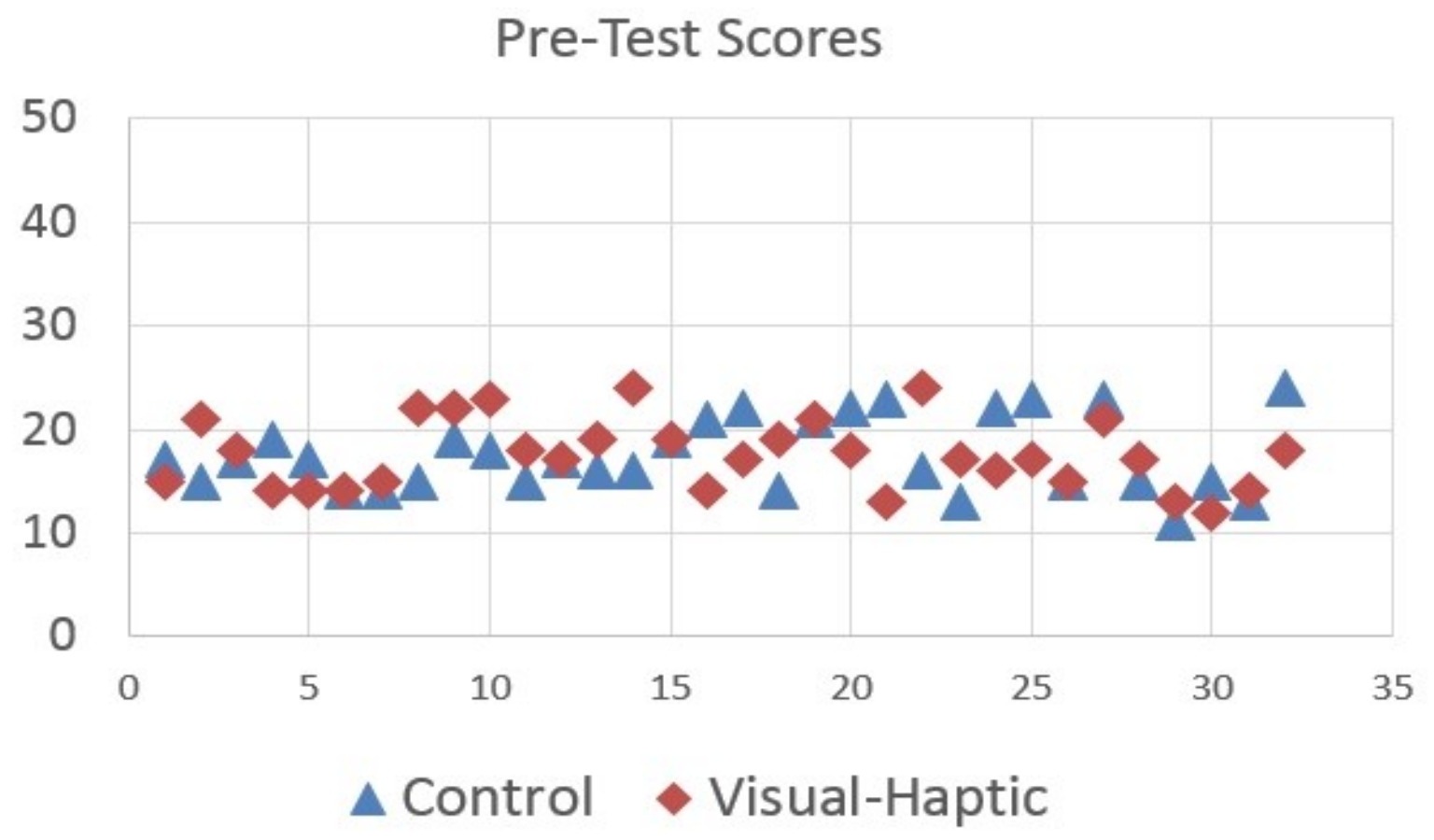

Figure 7. Pre-test scores for both, C and VH groups

The pre-test scores indicated that participants had little or no knowledge of the Gyroscopic Precession, with the average score for both groups being 17.35%, while a random chance trial would yield a score of 16%, as illustrated in Figure 7. Moreover, a t-test on the group scores, indicated that there was no significant difference between the groups in terms of their background knowledge. An *ANOVA* and a *two tail t-test* on the pre-test scores, for the C and VH groups, did not detect a substantial difference between the two groups ($t = 0.04$ with $p\text{-}value = 0.83$).

## 6.2 Lecture and Laboratory Experimentation

All participants were given a fifty minutes lecture on Gyroscopic Precession using traditional materials i.e., textbook information and associated 2D figures. After a short break:

- The students in the Control group (C) participated in an additional fifty minute traditional laboratory class, with the instructor solving different problems related to Gyroscopic Precession, whilst interacting with the students by asking and answering questions.
- The students in the Visual-Haptic group (VH) participated in a fifty minute session of Visual-Haptic activities to *feel* and *see* the relationship between the force vectors by "playing" with the visual-haptic simulator. The instructor asked students to modify the spinning wheel weight, radius and angular velocity and observe the orientation and magnitude of the counterforces generated in the system. While students were experimenting individually with the simulator, two assistants were helping with small troubleshooting tasks in order to direct students' focus on the concept understanding.

### 6.3 Learning Assessment – Post-Test

At the end of the experimental session all participants received a comprehensive test on concepts linked directly to gyroscopic precession. The fifty minutes post-test consisted of a set of 25 questions; 15 multiple choice questions and 10 essay type questions, each having an equal weight in the final grade. The paper-based test was taken by all students, in a classroom environment, with no calculator or haptic device allowed during the examination. Figure 8 provides an overview of the post-test results for each group.

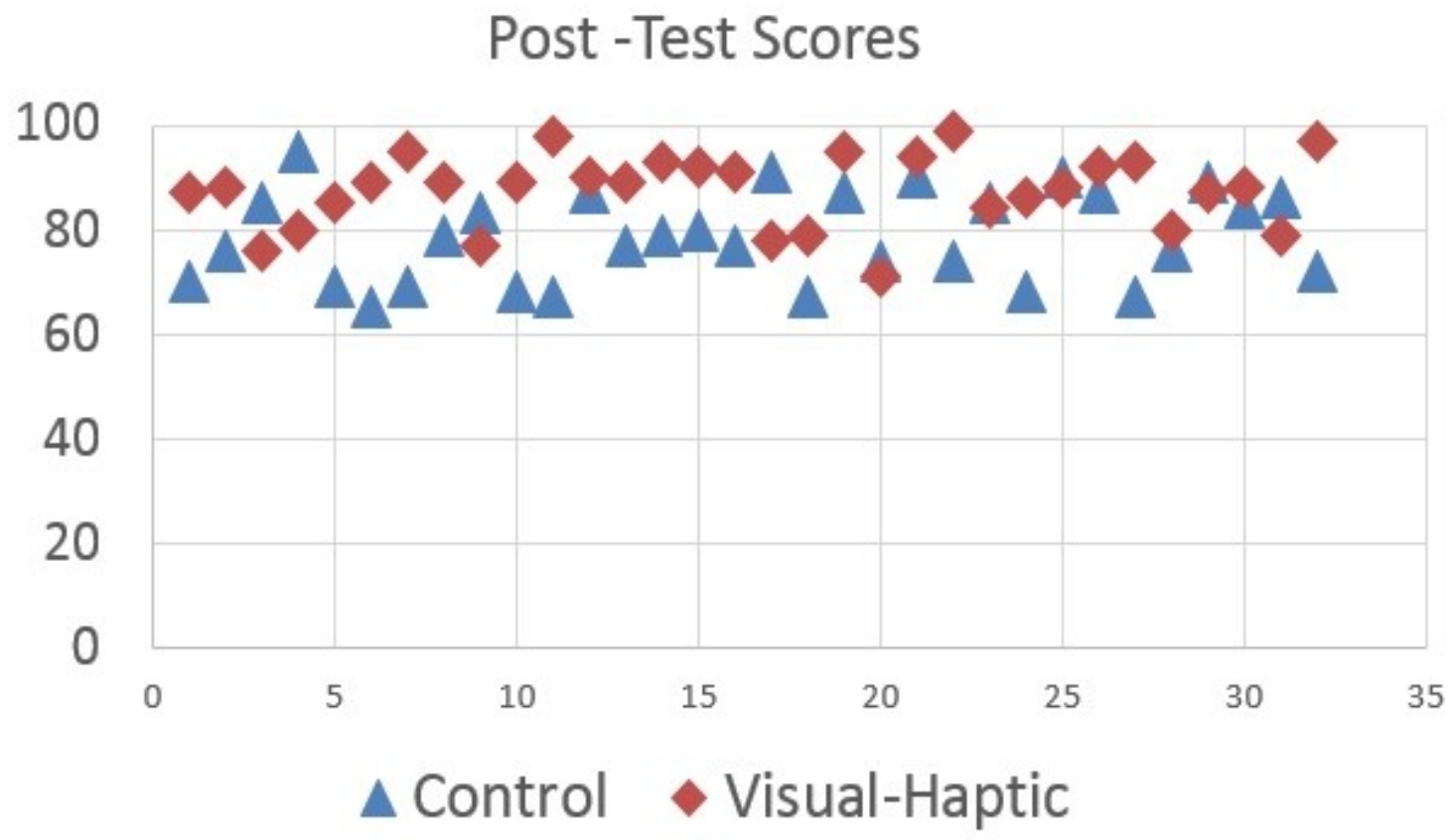

Figure 8. Post-test scores for both, C and VH groups

The post-test scores indicate, as expected, an increase in all the participants' knowledge of concepts associated with the Gyroscopic Precession, with an average overall score of *82.98%.*

### 6.4 Groups Performance Comparison

As an immediate observation, a much larger variance in the Post-test results exist, compared to the Pre-tests. The Pre-tests scores standard deviation (SD) for the control group, $C_{SD}=3.33$, while for the visual-haptic group, $VH_{SD}=3.23$. The Post-test SD for the control group was $C_{SD}=8.74$, while the visual-haptic group had a smaller value, $VH_{SD}=6.96$, as illustrated in Figure 9 (maximum score was 100), indicating a wider spread among participants' scores after instruction.

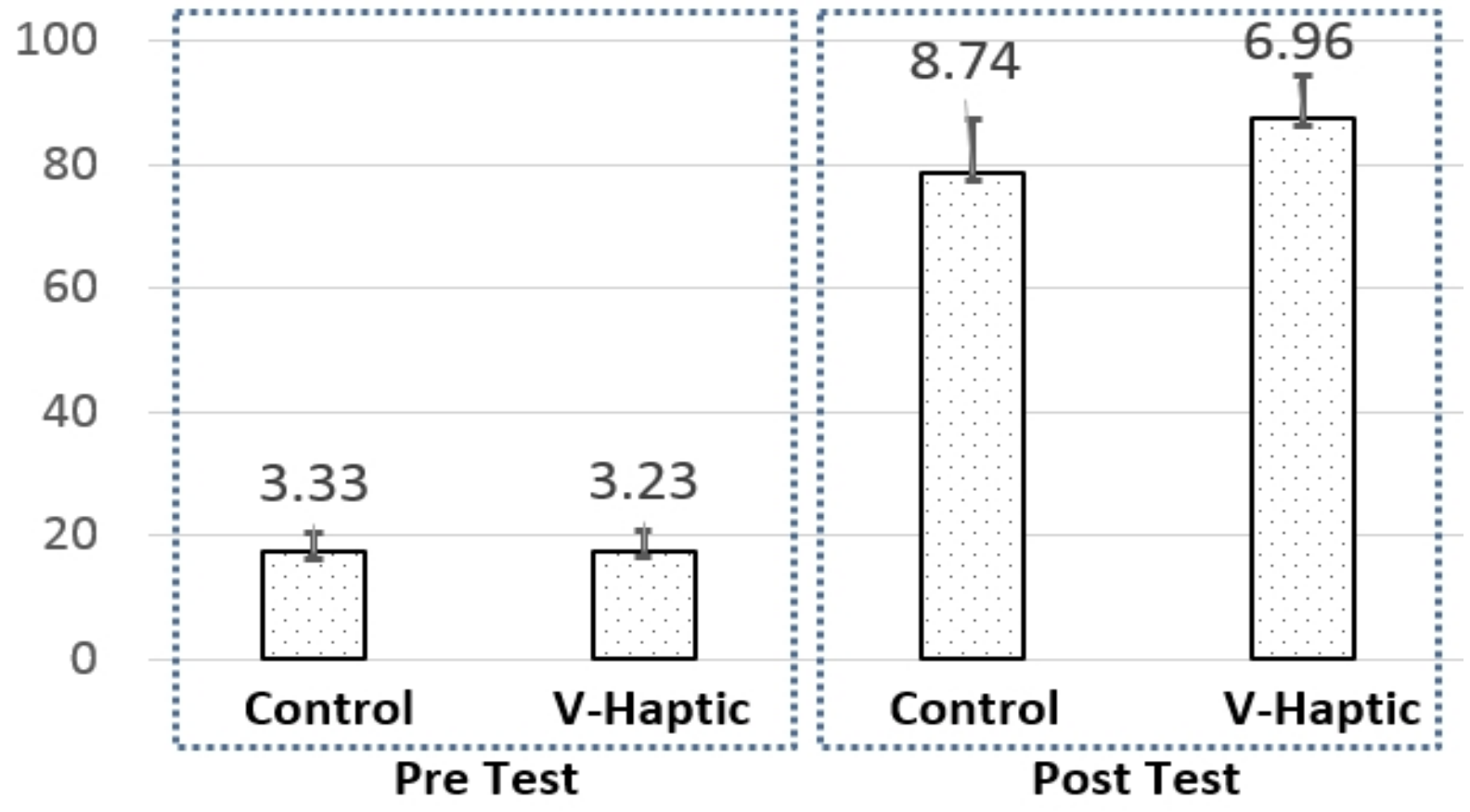

Figure 9. Pre and Post-Tests averages and standard deviations

As observed herein, the post-test score have a larger variance, while the students scored consistently higher in the VH group having the highest average (*87.44* out of *100* points) and the SD of *6.96*. The knowledge gain is clearly illustrated in Figure 9, as the overall scores significantly improved after lecture delivery and laboratory practice.

To further analyze the results, a *t-test* was computed on the post-test scores on the C and VH groups. The *t-test* proves a significant difference ($t=4.09$, $p < 0.001$ and $t_{critical}=1.69$) between the C and the VH groups. The participants in the VH group have consistently scored higher than the ones in the C group. The difference is an indicator to the potential improvement in learning as a result of employing the visual-haptic simulator for gyroscopic precession understanding. A histogram of the post-test results for each group shows a clustering of the VH group closer to the 100% mark for the grade, than the C group, whilst the C group has a wider distribution further from the maximum as illustrated in Figure 10.

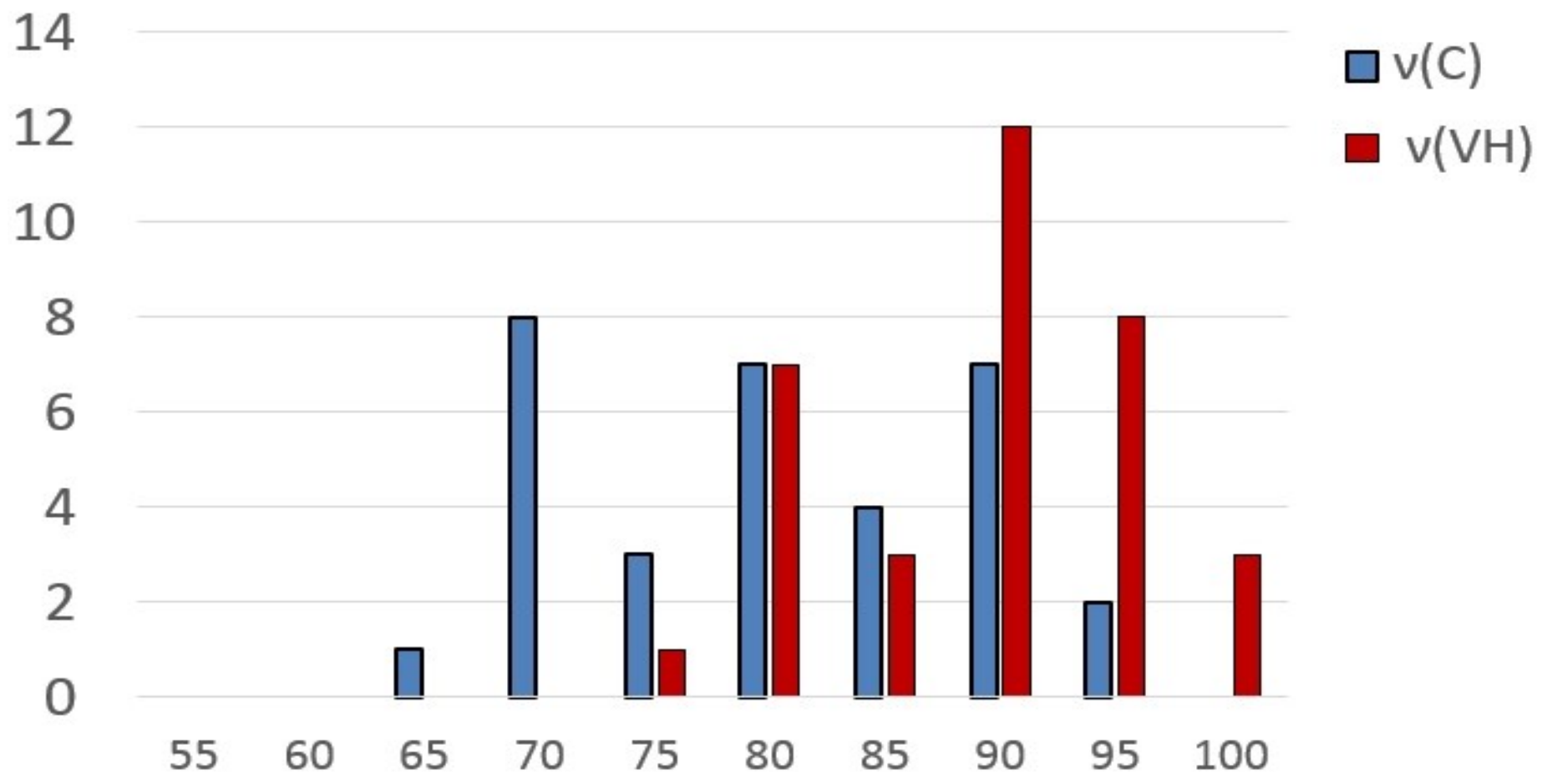

Figure 10. Histogram of the post-test scores for both groups.
Vertical axis - frequency, Horizontal axis - test score bins.

## Conclusion

Haptic technology is still in its infancy. We proposed an application of haptic (force-feedback) system in augmenting existing laboratory and lectures, to promote the understanding of the gyroscopic precession. The 3D visual environment with haptic (tactile) feedback, meets the needs of kinesthetic learners and could enhance their concept understanding. From early childhood, humans have been taught to learn using all of their senses, and one learns better when s/he can apply more senses to a specific task. Combining multiple modalities (visual, tactile, auditory etc.) may increase the learner's attention and retention of concepts.

While the results of our experiments have certain limitations (i.e., number of users, user's bias, other co-depended variables, etc.), visual-haptic simulators in education may offer an effective solution to time and cost restrictions in the very near future. With technology advancing at a fast pace, and haptic device prices declining rapidly, many new applications are being developed to take advantage of this new human-computer interaction modality. Our goal is to investigate the potential of visual-haptic simulators in enhancing learning, by providing access to revolutionary perspectives on abstract concepts, in ways that are not possible through the traditional method. We must continue to explore applications that provide revolutionary learning environments as the technology evolves. There are many abstract concepts that fit well into haptic implementations, but it is a challenge to define, design and implement such user interfaces. As it pertains to education, haptic interfaces can mainly address the experience portion of the learning cycle. Additional effort must be directed into encouraging and mediating a reflection phase in order to improve the learner's performance (Rose et al, 2018).

Like with any novel user interface, care must be taken to avoid overloading the user cognitive capacity. For this reason, a few days before the actual experiments, the students from the VH group were invited in the laboratory to play a "*space tennis*" game (a 3D haptic game scenario part of the H3D API) using the Falcon haptic devices. The purpose of this practice was to improve the students' familiarity with the haptic system, and to reduce the cognitive load during the actual experiments.

As of now, there are many constraints on effectively using haptics in education. The touch component in haptic devices is simulated through complex computations (in the discrete, *digital* domain), and at this time we can only experience a limited range of feelings (in our infinite, *analog* domain). Haptic devices are mechanical devices, and they inherently introduce variables such as friction among mechanical components, that are not related to the simulations. Moreover, the forces applied by the users can damage the haptic interface if a sufficiently large force is applied to the device. The virtual environment is comprised of pixels, and participants use visual cues to connect what they feel with what they see. Not all computers will have optimal resources to devote to the simulation, therefore, some visual lag may occur and cause disconnections between what the user feels and what the user sees. Nonetheless, the potential advantages of using haptics greatly outweighs the disadvantages, and as technology matures many of these problems will be addressed.